# The direct relation between the coefficient of the low temperature resistivity $T^2$ term and the superconducting transition temperature $T_C$

Manuel Núñez-Regueiro[1], Gastón Garbarino[1†] and María Dolores Núñez-Regueiro[2#]

1 Institut Néel, Centre National de la Recherche Scientifique (CNRS) & Université Joseph Fourier (UJF), 25 Avenue des Martyrs, F-38042 BP166 Grenoble Cedex 9 France

2 Laboratoire de Physique des Solides, Bât. 510, Université de Paris-Sud, Orsay, France

In several superconductors above the superconducting transition temperature $T_c$, the electrical resistivity is of the form $\rho = AT^2$. We show that there exists an empirical relation between $T_c$ and $A$ when both vary with an external parameter, e.g. pressure. The more resistive the sample the higher the $T_c$. Landau theory shows that it is a general feature of Fermi Liquids, as $\rho$ is governed by the scattering that bounds the pairs condensing at $T_c$. We develop a method that allows the determination of the coupling constant $\lambda$ that is validated when used to the transport properties of superfluid $^3He$.

---

† Present address : European Synchrotron Radiation Facility (ESRF), 6 Rue Jules Horowitz 38043 BP 220 Grenoble
# Deceased

An approach towards the behaviour of electrons in metals is the Landau theory of Fermi liquids, where electron interactions are accounted for by dressing the electrons with effective masses $m^*$ into new quasiparticles, with properties similar to those of the bare electrons. At low temperatures, this picture yields an electrical resistivity due to electron-electron scattering $\rho_{ee}$ that can be written $\rho_{ee} \propto m^*\tau^{-1} \propto \langle W \rangle m^{*2} T^2 \propto AT^2$, where $\langle W \rangle$ is the average scattering probability. As the specific heat of a FL is $C_v = \gamma T \propto m^* T$, the ratio $A/\gamma^2$ was calculated[1] for many materials and found effectively to be constant within different families (Kadowaki-Woods)[2]. In this paper we will focus on the scattering probability and its relation to superconductivity.

There are a significant number of superconductors from diverse venues that present the FL $AT^2$ dependence above the superconducting transition. For example, in heavy fermions both $A$ and $T_c$ are thought to be due to the same mechanism, i.e. spin fluctuations[3,4,5]. FL behaviour also appears in conventional high temperature superconductors, as in *A-15* superconducting compounds[6]. Or in low $T_c$ *Al* metal, where it was concluded that the electron-electron interaction was intermediated by phonons [7]. Successful efforts towards an empirical understanding normally compare different compounds, e.g. Kadowaki-Woods ratio. However, changing the properties of the same sample with an external parameter is often a more powerful method, as it does not change sample quality by means of uncontrolled variation of impurity concentration. A long term study of superconductors, together with the reanalysis of published data as well as new data presented in Fig. 1 a and b (obtained using the method described in Ref. 8), has allowed us to determine a direct relationship between the coefficient $A$ and the superconducting transition temperature, $T_c$, that we present on Fig. 1c. The empirical relationship that follows from the results shown on this figure has never been reported (nor even addressed). It scans three orders of magnitude of $T_c$ and several of resistivity. From Fig 1c $T_c$ is a monotonous increasing function of $A$, the stronger the scattering the higher the $T_c$. In simple words, it is a clear quantitative manifestation of the well-known thumb rule: the worst metals give the best superconductors.

We can understand this relation within Landau theory by analyzing both the quasiparticle scattering and the superconducting transition temperature.

We reduce the inverse quasiparticle-quasiparticle scattering time and the transition temperature to the superfluid phase, to the Landau amplitude scatterings within the *s-p* approximation. From standard Landau FL theory[9] we obtain the inverse quasiparticle-quasiparticle scattering time $\tau^{-1}$,

$$\tau^{-1} = \frac{m^{*3}<W(\theta,\phi)>(k_BT)^2}{8\pi^4\hbar^6} = \frac{m^*<W(\theta,\phi)>N(k_F)^2(k_BT)^2}{8\pi^2\hbar^2V^2E_F} \quad (1)$$

Where $m^*$ is the effective mass, $V$ the volume, $N(k_F)$ the density of states at the Fermi level, $<W(\theta,\phi)>$ is the transition probability which describes the scattering of two quasiparticles whose momenta are related by the standard angles $\theta$ and $\phi$, the bracket indicating the angle average through the Abrikosov-Khalatnikov angles; $k_B$ the Boltzmann constant and $\hbar$ the Planck's constant. Developing in Legendre polynomials within the *s-p* approximation [10] we obtain for the triplet transition probability

$$W_{\uparrow\uparrow}(\theta,\phi) = \frac{2\pi}{\hbar}\left|\frac{A_{\uparrow\uparrow}(\theta,\phi)}{N(k_F)}\right|^2 \sim \left|\frac{(1+A_0^a)}{N(k_F)}\right|^2 \quad (2)$$

and

$$(\tau \cdot T^2)^{-1} \sim \frac{m^*k_B^2(1+A_0^a)^2}{4\pi\hbar^3V^2E_F} \propto (1+A_0^a)^2 \quad (3)$$

where $A_{\uparrow\uparrow}(\theta,\phi)$ is the triplet scattering amplitude and $T_F$ the Fermi temperature and $A_l^\sigma$ the Landau scattering amplitudes (we have taken $A_0^s=1$). A similar result, $(\tau \cdot T^2)^{-1} \propto (2A_1^{ep})^2$, is obtained for singlet scattering taking $A_l^s = A_l^a = A_l^{ep}$, as in this case the interaction does not depend on spin direction.

Now, following Ref.[11], Patton and Zaringhalam[12] (PZ) estimated the transition temperature to a condensed state as a function of the Landau scattering amplitudes, $T_F$ being the Fermi temperature and $\lambda_{\uparrow\uparrow}$ the triplet coupling constant, and $\alpha$ a parameter,

$$T_c = 1.13\alpha T_F e^{1/\lambda_{\uparrow\uparrow}}$$

$$\lambda_{\uparrow\uparrow} = A_{\uparrow\uparrow}(\pi,\phi)/3\cos(\phi) = \sum_l (-)^l (A_l^s + A_l^a)/12 \approx -(1+A_0^a)/6 \quad (4)$$

where we have limited the development to $l = 0,1$ within the *s-p* approximation and we have neglected singlet scattering. For singlet scattering, neglecting triplet scattering, we have $\lambda_{\uparrow\downarrow} \approx 2A_1^{ep}$. From (3) and (4) we conclude that $(\tau T^2)^{-1} \propto (\lambda_{\uparrow\uparrow})^2$, the same for singlet scattering. Thus

$$T_c = \theta e^{\varsigma/\sqrt{(\tau T^2)^{-1}}} \quad (5)$$

where $z$ and $q$ are parameters. We find that both $\langle W \rangle$ and the coupling parameter $l$, that defines $T_c \propto e^{-1/\lambda}$, depend on the same combination of scattering amplitudes, yielding $\lambda^2 \propto \langle W \rangle \propto (\tau T^2)^{-1}$. In other words the same scattering that controls the transport is the one responsible for the superconducting pairing.

The physics of FL should be general. Therefore and before continuing our analysis on metals it is advisable to test this result in a different, well studied, FL. Consider the case of liquid $^3He$, where the Fermions are the interacting $^3He$ atoms. It displays a transition at milikelvin temperatures to a superfluid state, due to a BCS condensation of atom pairs. As liquid $^3He$ is chargeless, there is no electrical resistance to analyze. However, there are three transport properties that also depend on $\tau \cdot T^2$: the thermal conductivity ($\kappa$), the spin diffusion (*SD*) and the viscosity ($\eta$). To apply expression (5), we must consider that for $^3He$, $\theta = 1.13\alpha T_F^{**}$, $T_F^{**}(P)$ being the Fermi temperature renormalized by all the interactions at each pressure. To extract the pressure dependence from $\theta$, we normalize $T_c(P)/T_F^{**}(P)$, and fit it using expression (5) to the measured $\tau \cdot T^2$ as the pressure $P$ is varied with $\alpha$ as a parameter instead of $\theta$(Figures 2a to c). We can thus extract $\alpha \approx 0.2$ and $\varsigma$, and hence the coupling constant $\lambda = \sqrt{(\tau T^2)^{-1}}/\varsigma$. Fig. 2d compares the superfluid coupling constant $\lambda$ for different pressures obtained from the three fits and those calculated theoretically[36]. The agreement is excellent, although we must consider that all pressure effects other than those on $\tau$ are probably taken into account by the renormalization with $T_F^{**}(P)$. The application on $^3He$ confirms our analysis based on the idea that the pre-transitional scattering determines $T_c$.

Having shown the utility of expression (5), we come back to the superconductors of Fig. 1c. The FL resistivity of metals[13] is given by $\rho_{ee} = \dfrac{m^*}{ne^2 \tau_R}$, where n is the carrier density, $m^*$ the effective mass, $e$ the electronic charge and $\tau_R = number \times \tau$. From (1), (2) and (4) we find now that

$$\rho_{ee} \approx \frac{m^{*2} k_B^{\,2} (1+A_0^a)^2}{ne^2 \, 4\pi \hbar^3 V^2 E_F} T^2 \qquad\qquad A \approx \frac{m^{*2} k_B^{\,2}}{ne^2 \, 4\pi \hbar^3 V^2 E_F} \lambda_{\uparrow\uparrow}^2 \qquad (6)$$

A similar result can be obtained for singlet scattering. As $A \propto (\tau T^2)^{-1}$, we can now attempt an equivalent type of fit $T_c = \theta e^{-\varsigma/\sqrt{A}}$ (7). We must bear in mind, though, that application of elementary FL to materials with complex Fermi surfaces is bound to be cumbersome. In this case other parameters that are now present besides $\lambda$ in $A$, may vary, as well. Also, we ignore the pressure dependence of the $\theta$ parameter ($\theta \propto 1.13\hbar\omega_D$ for conventional superconductors, where $\omega_D$ is the Debye frequency). We show on Fig. 3a an example of a fit on the $Nb_3Sn$ data, that allows us to obtain $\theta$ and $\varsigma$, and hence the coupling constant $\lambda = \sqrt{A}/\varsigma$. It yields $\lambda$ =2.25 at ambient conditions, which is similar to that obtained from other methods ($\lambda$=2.3[14], 1.7-2.0[15]). Our weak coupling approach gives a reasonable accord for the strong coupling material $Nb_3Sn$. This comes probably from the fact that, neglecting the Coulomb electron-electron interaction ($\mu^* \approx 0.1$), the renormalization due to strong coupling ($\lambda \rightarrow \lambda/[\lambda+1]$[16]), is factorized out into our parameter $\theta \rightarrow \theta/2.7183$, from which no information is extracted. Thus, the variation of $T_c$ and $A$ seems to be controlled mainly by the variation of $\tau$.

We apply the method to all the compounds of Fig. 1c. For borocarbides we obtain a strong coupling (weak coupling) value for $YPdBC$ ($PrPt_2B_2C$) as expected, while for $YNi_2B_2C$ we obtain at ambient pressure $\lambda$=0.9, to be compared with $\lambda$=0.84 extracted from $H_{c2}$ measurements [17]. For $V_3Si$ we extract $\lambda$=0.87 in excellent agreement with the value, $\lambda$=0.89, from tunneling measurements[18]. The $MgB_2$ data render a low coupling $\lambda \approx 0.25$, due to the fact that the weak coupling 3D carriers control the resistivity. In the organic compounds the higher values ($\lambda$=0.37 from our fit) for $(TMTSF)_2ClO_4$ with respect to $(TMTSF)_2PF_6$ ($\lambda$=0.22 from our fit) follow the expected trend. While the values we obtain for the heavy fermion compounds are one order of magnitude lower than previously reported ones. The failure of the method for heavy fermions may be due to the fact that the variation of $A$ with pressure is controlled by $m^*$ and not by $\tau$. As our analysis works properly on irradiated $Nb_3Sn$, it may be interesting to study the variation of $T_c$ and $A$ in heavy fermions by using irradiation, not pressure, as defects should have a stronger effect on the scattering rate $\tau$ than on $m^*$.

Thus, with the exception of heavy fermions, the $\lambda$-s that we extract are in good agreement with those obtained by other methods, in spite of the crudeness of the FL model that we have used. For On Fig.3b we present the obtained data in one single graph, normalizing $T_c$ and $\sqrt{A}$ by the obtained $\theta$ and $\varsigma$ ($\lambda = \sqrt{A}/\varsigma$), respectively. We must note that the fit is very stable, as it is related to the simple property $\frac{d \ln T_c}{dP} = \frac{1}{2\lambda} \frac{d \ln A}{dP}$, that can be derived easily from (7). All the dispersed data of Fig1c collapse onto one single curve, that scales the compounds according to their respective superconducting coupling constants, derived from their resistivity coefficients *A*.

Extrinsic origins of quadratic temperature terms can also be present[19] and should be taken into account. This is the case of inelastic scattering by impurities or Koshino-Taylor effect (KT). We have earlier shown[20] that KT is at the origin of the quadratic term of the superconductor *NbTi* , where *A* increases with pressure while $T_c$ decreases. Intermediate cases can happen, where both terms are important but even in that case the correct $\lambda$ may be obtained[21] through corrections (subtraction of a KT term of the form $A_{KT} \approx 10^{-5} R_0$ , where $R_0$ is the residual resistivity), if there is a correct correlation between *A* and $T_c$. Clearly, further theoretical work is needed to understand the interaction of the FL scattering rate with the defect induced KT term[6].

In metals, besides the cases described above, $T^2$ terms have been also reported in alkali metals, where it is attributed to e-e Coulomb interaction[7]. It is though unobservable in elemental metals and many superconductors where the Bloch-Grüneisen type dependences ($T^n + T$ ; $\sim 3 \leq n \leq 5$) due to electron-phonon scattering predominate. In order to show a $T^2$ term, a strong effective e-e phonon mediated interaction [7], that will eventually lead to pairing, must be present. The stronger the phonon mediated e-e interaction, the shorter the scattering time, the higher the resistivity and the stronger the pairing. However, too strong scattering invalidates the quasiparticle picture, causing the breakdown of Landau theory. As a consequence, a marginal Fermi liquid type temperature dependence ($\rho \propto T$) appears. This situation is of actual interest in, i.e. high temperature superconducting cuprates[22], although here the intermediating boson is not necessarily a phonon.

In conclusion, we have shown that there is a direct empirical relation between the coefficient *A* of the quadratic in temperature FL low temperature resistivity term and the superconducting transition temperature $T_C$ for a large number of superconductors. The empirical relation can be understood within Landau theory of FL. Finally, we formulate accordingly a method to obtain the superconducting coupling constant $\lambda$ from the evolution of both variables under an external parameter. This method is validated by its application on the evolution with pressure of the transport properties and the superfluid transition temperature of liquid helium 3.

**Acknowledgements**

Application to $^3He$ owes much to E. Collin and H. Godfrin. M.N-R acknowledges discussions with J. Ranninger, late L. Puech, G. Vermeulen, C. Balseiro, A. Aligia, J-P. Brison and P. Nozières; also a careful reading of the manuscript by J. E. Lorenzo, P. Monceau and M. A. Continentino. This work was partially supported by the project TetraFer ANR-09-BLAN-0211 of the Agence Nationale de la Recherche of France.

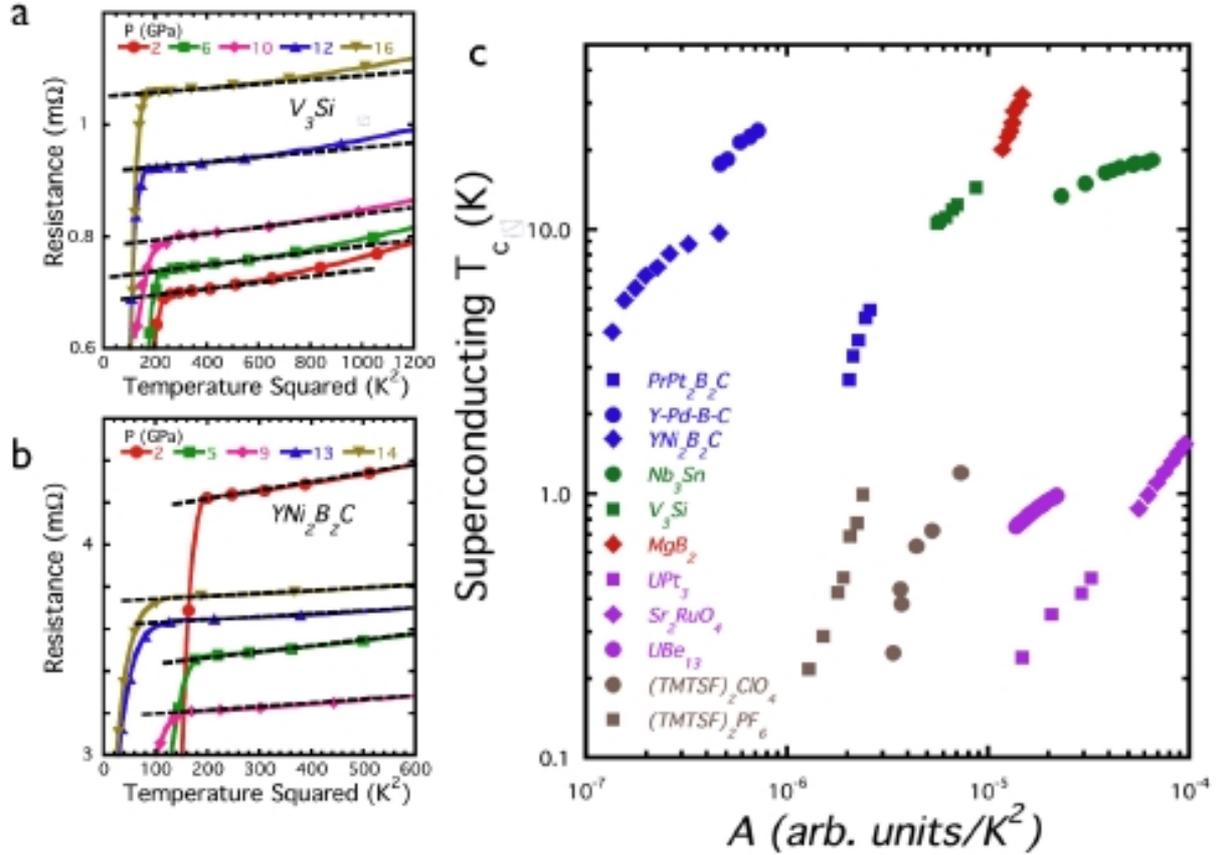

**Figure 1**

(a) The electrical resistivity of $V_3Si$ for different pressures as a function of $T^2$ showing the quadratic behavior some degrees around 22K as reported previously[23] ; this range increases with pressure (only one sixth of the points are shown for clarity). (b) The electrical resistivity of $YNi_2B_2C$ for different pressures as a function of $T^2$ showing the quadratic behavior (only one sixth of the points are shown for clarity). (c) Empirical relation between $T_c$ and the coefficient $A$ of the quadratic temperature term for different values of an external parameter is applied, pressure unless specified otherwise. $PrPt_2B_2C$(blue squares) [24]; Y-Pd-B-C(blue dots) [25]; $YNi_2B_2C$(blue diamonds); $Nb_3Sn$ ($\alpha$ or $e^-$ irradiation, green dots)[23]; $V_3Si$ (green squares); $MgB_2$(red diamonds) [26]; $UPt_3$(magenta squares) [27]; $Sr_2RuO_4$(magenta diamonds) [28]; $UBe_{13}$(magenta dots) [29],[30]; $(TMTSF)_2ClO_4$ (brown dots) [31], $(TMTSF)_2PF_6$ (brown squares)[31].

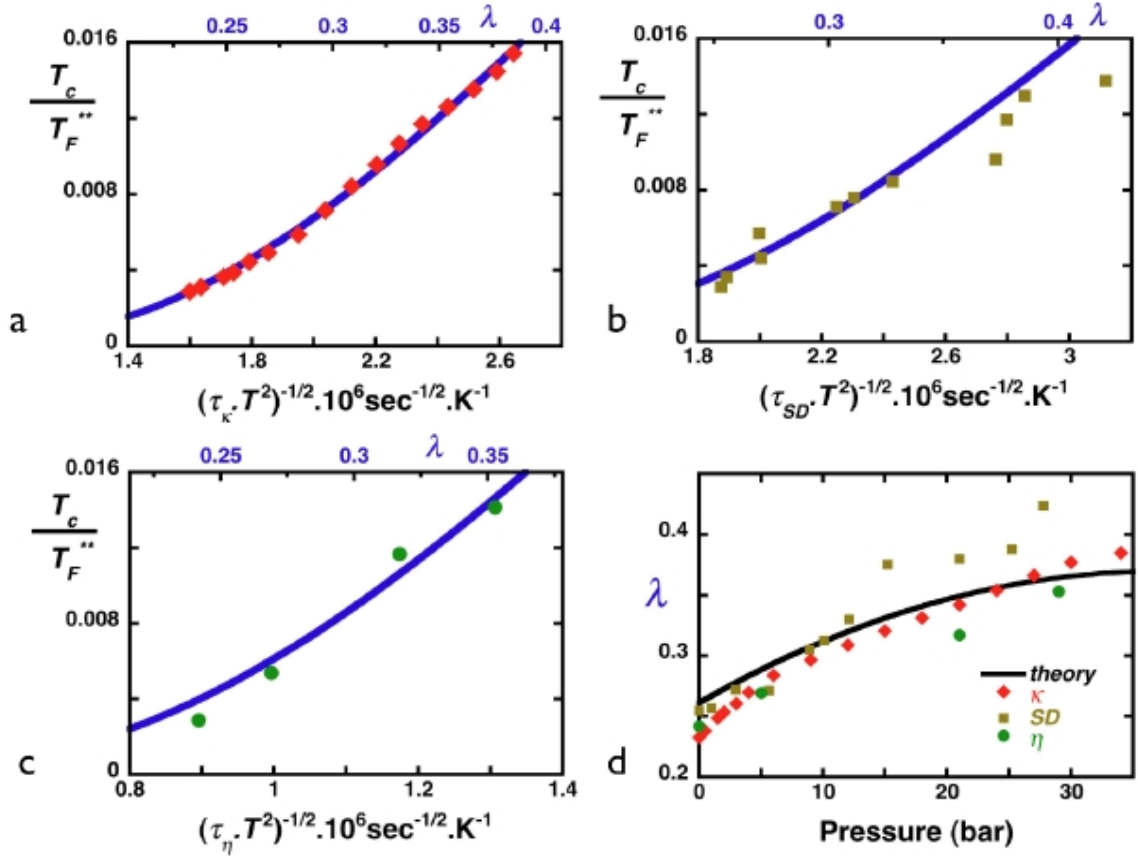

Figure 2 (a) Superfluid transition temperature $T_c$ of $^3He$ normalized by $T_F^{**}(P)$, the Fermi temperature renormalized by all the interactions at each pressure[32], as a function of the inverse square root of the scattering time extracted from the thermal conductivity[33] multiplied by the square of the temperature. Fitting with $T_c/T_F^{**} = 1.13.\alpha e^{-\varsigma/\sqrt{(\tau T^2)^{-1}}}$ (blue line) allows the determination of the parameter $\varsigma$ and the coupling constant $\lambda = \sqrt{(\tau T^2)^{-1}}/\varsigma$. (b) Same treatment applied to the spin diffusion (*SD*) scattering (data from ref. 34) (c) Same treatment applied to the viscosity ($\eta$) scattering rate (data from ref. 35). (d) Comparison of the superfluid coupling constant $\lambda$ values obtained from our fits to the ones obtained theoretically (black solid line)[36]. The agreement between theory and our analysis for the tree transport properties is excellent, validating the analysis, considering that we have neglected the corrections due to angle averages in the scattering integral, that are different for each property.

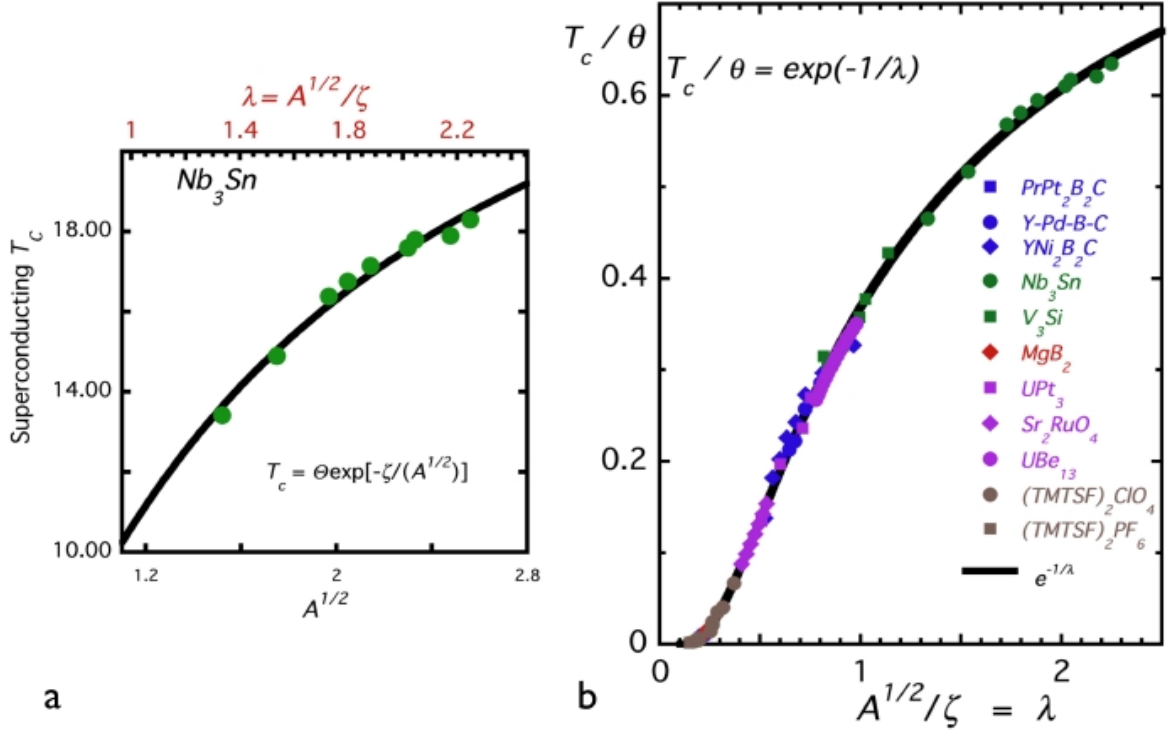

**Figure 3**

(a) Fit of the weak coupling expression $T_c = \theta \cdot e^{-\varsigma/\sqrt{A}}$ assuming the approximation that $\Theta$ does not vary with the external parameter ($\alpha$ or $e^-$ irradiation) and that all the variation of $A$ is due to the variation of the scattering rate $\tau$. Neglecting the variation of $\theta$, we obtain $\lambda = \sqrt{A}/\varsigma \sim 2.2$, that in very good agreement with previously reported values (see text). (b) The same type of fit shown for the compounds of Fig. 1, presented now with each $T_c$ normalized to the fitting parameter $\Theta$ and as a function of $\lambda = \sqrt{A}/\varsigma$. The values of $\lambda$ we obtain for *A-15* compounds and borocarbides agree within 10% with those obtained from other methods, while those of heavy fermions are much lower than reported previously (see supplementary material for a detailed discussion).